# Extraordinary stiffness tunablility through thermal expansion of nonlinear defect modes

Marc Serra Garcia*[1], Joseph Lydon[2] and Chiara Daraio[1,2]

[1]Department of Mechanical and Process Engineering, Swiss Federal Institute of Technology (ETH), Zürich, Switzerland

[2]Engineering and Applied Science, California Institute of Technology, Pasadena, California 91125, USA

*email: sermarc@ethz.ch

**Incremental stiffness characterizes the variation of a material's force response to a small deformation change. Typically materials have an incremental stiffness that is fixed and positive, but recent technologies, such as super-lenses[1], low frequency band-gap materials[2] and acoustic cloaks[3,4], are based on materials with zero, negative or extremely high incremental stiffness. So far, demonstrations of this behavior have been limited either to a narrow range of frequencies[2], temperatures[5], stiffness[6] or to specific deformations[7,8]. Here we demonstrate a mechanism to tune the static incremental stiffness that overcomes those limitations. This tunability is achieved by driving a nonlinear defect mode in a lattice. As in thermal expansion, the defect's vibration amplitude affects the force at the boundary, hence the lattice's stiffness. By using the high sensitivities of nonlinear systems near bifurcation points, we tune the magnitude of the incremental stiffness over a wide range: from positive, to zero, to arbitrarily negative values. The particular deformation where the incremental stiffness is modified can be arbitrarily selected varying the defect's driving frequency. We demonstrate this experimentally in a compressed array of spheres and propose a general theoretical model. This approach opens a new paradigm to the creation of fully programmable materials.**

Research on materials with mechanical properties not found in natural systems is a very active field. This research effort has resulted in different solutions: metamaterials[9], materials undergoing phase transitions[5,10,11] or buckling instabilities[7], materials with electromagnetic coupling between constituents[6,12]. Metamaterials are periodic systems with local resonances that can present negative or zero effective mass and stiffness[2, 9, 13, 14], however, their practical application is limited to a narrow band around the local resonance frequency[9]. Materials that operate around phase transitions[10, 15], or buckling instabilities[7] can achieve extreme negative values of the incremental stiffness, but their operation is limited to the deformation or temperature where the instability occurs[5]. Negative stiffness inclusions in a matrix, used to achieve positive extreme stiffness, suffer from stability problems[16]. Finally, many of the proposed solutions are limited in the range of attainable stiffness [6].

In order to address those limitations, we demonstrate tuning the incremental stiffness of a lattice using a method based on the thermal expansion of defect modes. Thermal expansion is an ubiquitous property of anharmonic lattices[17], in which the lattice can be made to expand or contract by increasing or decreasing its vibrational energy. In our method, we drive a defect mode in a lattice with a harmonic signal. As a consequence of anharmonicity in the lattice, an external deformation moves the defect mode in and out of resonance, affecting its vibrational amplitude. These changes in vibrational amplitude affect the thermal expansion of the defect, and therefore the force at the boundary. This alters the incremental stiffness of the lattice. We use this concept to achieve negative stiffness (Fig. 1a).

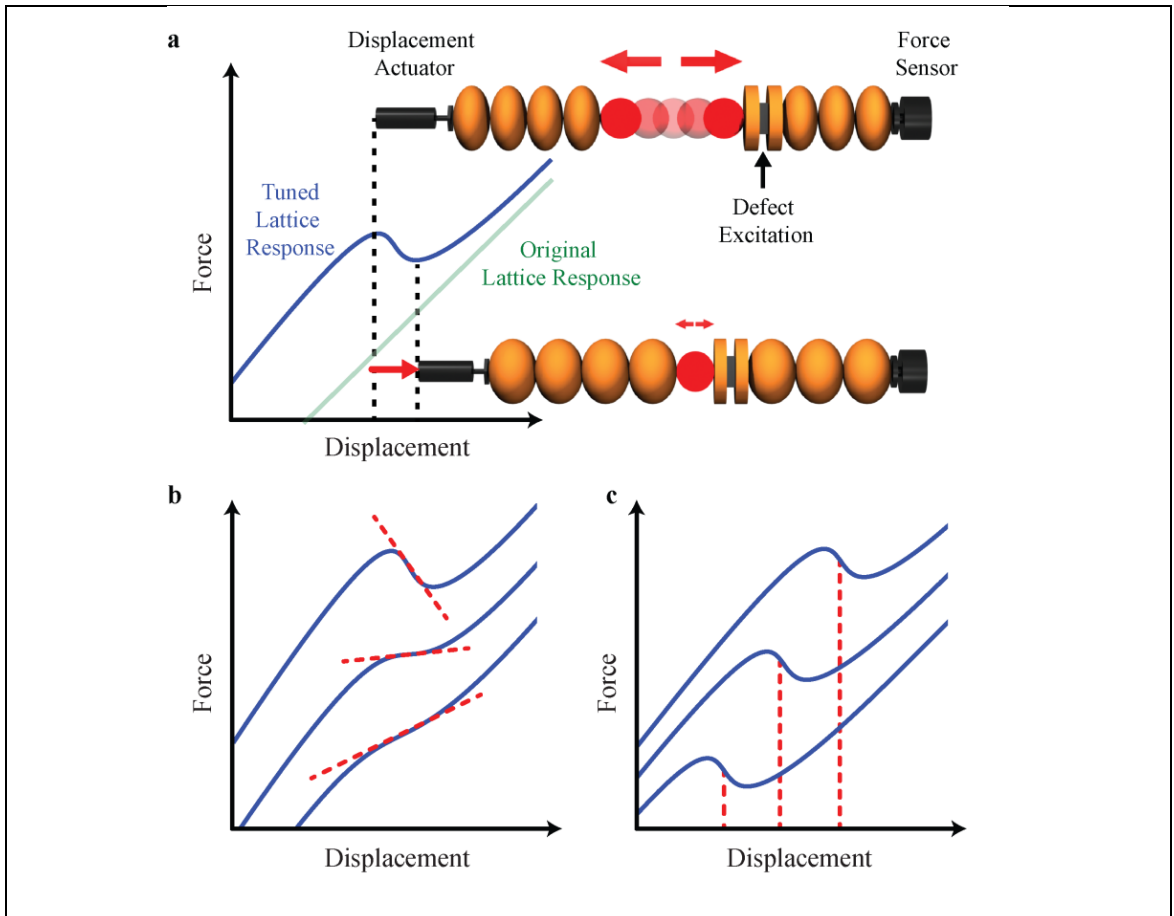

**Figure 1 | Tuning stiffness through thermal expansion. a.** Schematic diagram of the tunable stiffness mechanism illustrated in a 1-D granular chain. The diagram shows the static force on the lattice due to a prescribed displacement while harmonically driving the defect mode. As the lattice is compressed (blue arrow), the defect vibrational amplitude decreases (red arrows). This results in a negative incremental stiffness due to thermal contraction of the defect mode. **b.** Changes of the driving frequency and amplitude of the excitation control the incremental stiffness, and **c.** the strain point at which the stiffness is being modified. The curves are offset for clarity.

We demonstrate the concept in a one-dimensional lattice of coupled spheres (see Fig. 1a, Methods and Supplementary Information). The spheres interact through a nonlinear

Hertzian contact [18]. The central particle is a defect that allows the existence of a localized vibrational mode[18, 19, 20]. A piezoelectric actuator is embedded in the vicinity of the defect particle and is used to harmonically excite the defect mode. We monitor the defect mode vibration using a laser vibrometer. We acquire the quasi-static force-displacement relation of the lattice, by prescribing an external deformation and measuring the force at the opposite boundary. The vibration of the defect mode affects the force-displacement relation. The amplitude and frequency of the defect excitation are the two control variables that determine the mechanical response. Using these variables we can select both the incremental stiffness magnitude (Fig. 1b and Supplementary Video 1) and the displacement point where the incremental stiffness is being modified (Fig. 1c and Supplementary Video 2). This allows tuning the force-displacement response of a lattice at a selectable displacement value, a capability that exists so far only in biological organisms[21].

In our system the measured force depends on both the applied displacement and on the amplitude of the mode, $F(x, A)$. Therefore, the incremental stiffness, defined as the total derivative of the force with respect to the displacement, is given by the equation:

$$\frac{dF}{dx} = \left(\frac{\partial F}{\partial x}\right)_A + \left(\frac{\partial F}{\partial A}\right)_x \frac{\partial A}{\partial x}$$

The first term on the right side of the equation gives the stiffness of the lattice neglecting any change in the defect mode's amplitude. The second term describes the effect of the oscillation of the defect mode. The function $(\partial F/\partial A)_x$ is the change in the force due to a change in amplitude of the defect mode and quantifies the intensity of the thermal expansion. From a dynamical point of view, this arises due to an asymmetry of the interaction potential[17] and in our lattice is always positive (see Supplementary Materials). Finally, the effect of the strain on the amplitude of the mode is contained in the quantity $\partial A/\partial x$.

The vibration amplitude's dependence on strain is a consequence of the harmonic excitation and of the nonlinearity present in the chain. The harmonic excitation results in a defect mode resonance, which occurs when the defect mode's frequency matches the excitation frequency. The nonlinearity relates the mode's frequency, $\omega_d$, to the lattice strain, $\delta_0$.[18] In our system the Hertzian contact results in the relationship, $\omega_d \propto x_0^{1/4}$. As a result, straining the lattice causes a change in the mode's frequency (Fig. 2a). If the mode's frequency approaches the excitation frequency, the mode gets closer to resonance, and therefore the oscillation amplitude increases. Conversely, if the mode frequency moves away from the excitation frequency, the oscillation amplitude decreases (Fig 2b.). This strain controlled resonance results in a dependence of amplitude on strain and therefore, in a non-zero $\partial A/\partial x$.

Different excitation frequencies cause the resonance to happen at different strain values (Figs 2a,b). This is due to aforementioned frequency strain relationship, which associates a particular resonance strain to each excitation frequency. By choosing the excitation frequency we are able to set the displacement region where the system is in resonance and the stiffness is being modified (Fig. 2b).

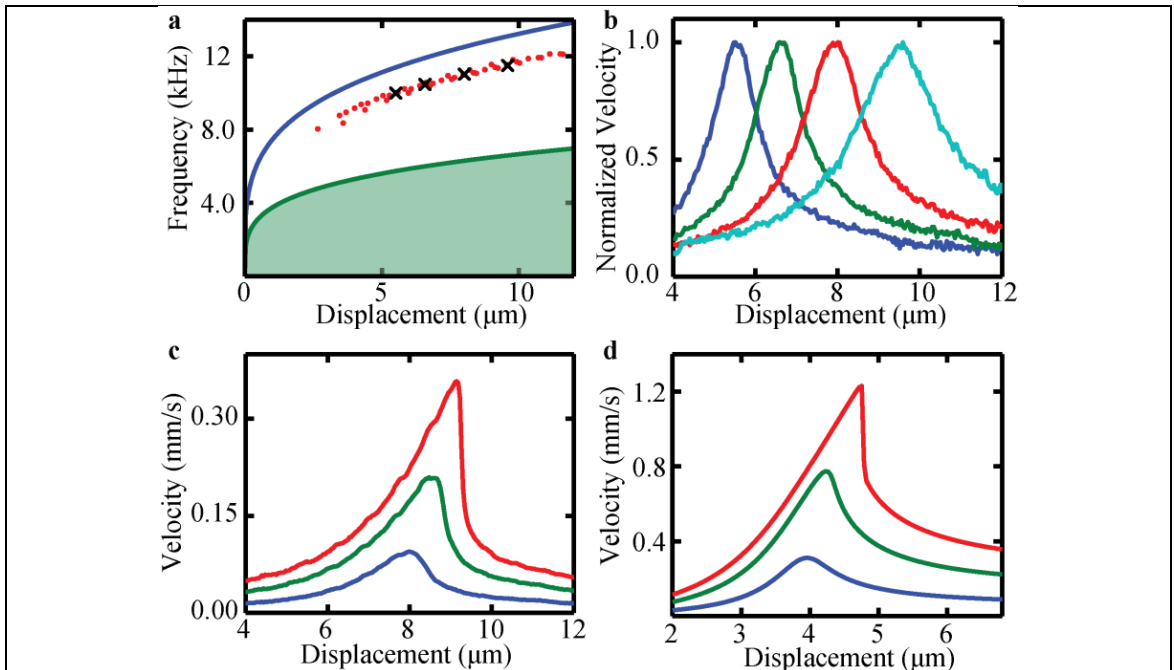

**Figure 2 | Response of the nonlinear defect mode. a.** Theoretical defect mode (blue) and acoustic band (green) frequencies dependence on prescribed displacement. Experimental measurements are plotted as red dots with the four curves in panel **b** marked with black crosses. **b.** Normalized experimental velocity of the defect mode as a function of displacement of the lattice. Curves correspond to excitation frequencies of 10(blue), 10.5(green), 11(red) and 11.5 kHz (cyan). **c.** Experimental velocity of the defect mode for drive amplitudes of 4.2 (blue), 9.8 (green) and 15.4 nm (red) all at 10.5 kHz. **d.** Numerical results corresponding to **c**, for defects driven at 20, 50, and 80 nm, respectively. Our simplified model (see Methods) qualitatively reproduces the experimental results, but is unable to make precise quantitative predictions.

The effect of the excitation amplitude on the defect's vibration is shown in Fig. 2c,d. As expected, driving the defect with larger harmonic forces results in larger oscillations. Furthermore, as the excitation amplitude gets larger the resonance response becomes increasingly asymmetric. This is a common property of driven nonlinear oscillators close to a bifurcation[22]. As nonlinear system's approach bifurcation points, oscillations become extremely sensitive to the strain[23]; in our system the magnitude of $\partial A/\partial x$ approaches minus infinity. This allows us to achieve arbitrarily large negative values of incremental stiffness.

These extreme negative values have been attained experimentally. The measured force-displacement curves at four different drive amplitudes are shown in Fig. 3. The incremental stiffness at our selected strain progressively decreases as the defect excitation is increased (Fig. 3a-d). For the largest excitation amplitude, the incremental stiffness reaches negative infinity (Fig. 3d). In order to validate that this effect is due to the defect's vibration, we simultaneously measure the defect's mode amplitude, presented below each force-displacement curve in Fig. 3a-d. The greatest change in the incremental stiffness happens where the slope, $\partial A/\partial x$, is the most negative. This occurs because larger changes in vibrational amplitude are accompanied by larger changes in thermal expansion. It should be noted that the negative stiffness values are stable because our experiment is done under prescribed displacement boundary conditions.

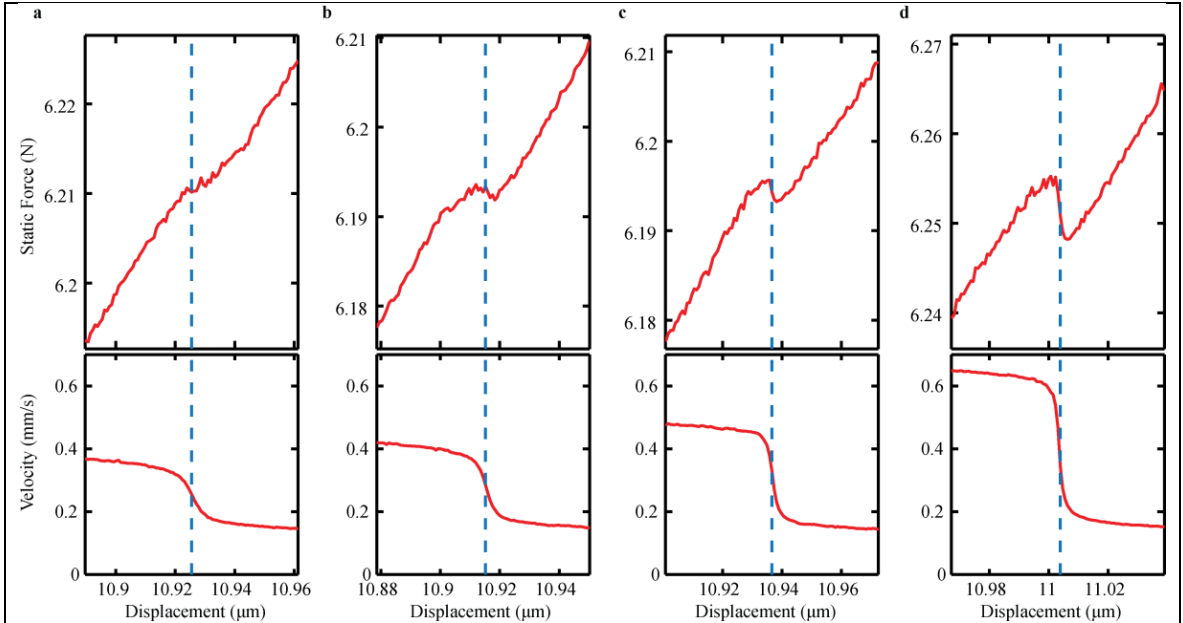

**Figure 3 | Experimental tuning of the incremental stiffness**. Force-displacement curves for excitation amplitudes of **a.** 5.9 nm **b.** 6.4 nm **c.** 7.54 nm **d.** 10.9 nm. Shown below are the defect mode velocities (proportional to the mode amplitude, $A$(x)) as a function of the overall displacement, $x$, of the lattice.

Each pair of drive frequency and amplitude results in a determined incremental stiffness at a particular displacement point. We explore this relationship analytically (see Supplementary Information) in Fig. 4a. The blue lines show contours at the same excitation amplitude and the red lines at the same frequency. To get a particular stiffness at a desired displacement, we select the excitation parameters corresponding to the lines passing through this point. While we only show a finite number of constant lines, all possible values in the shaded region are attainable.

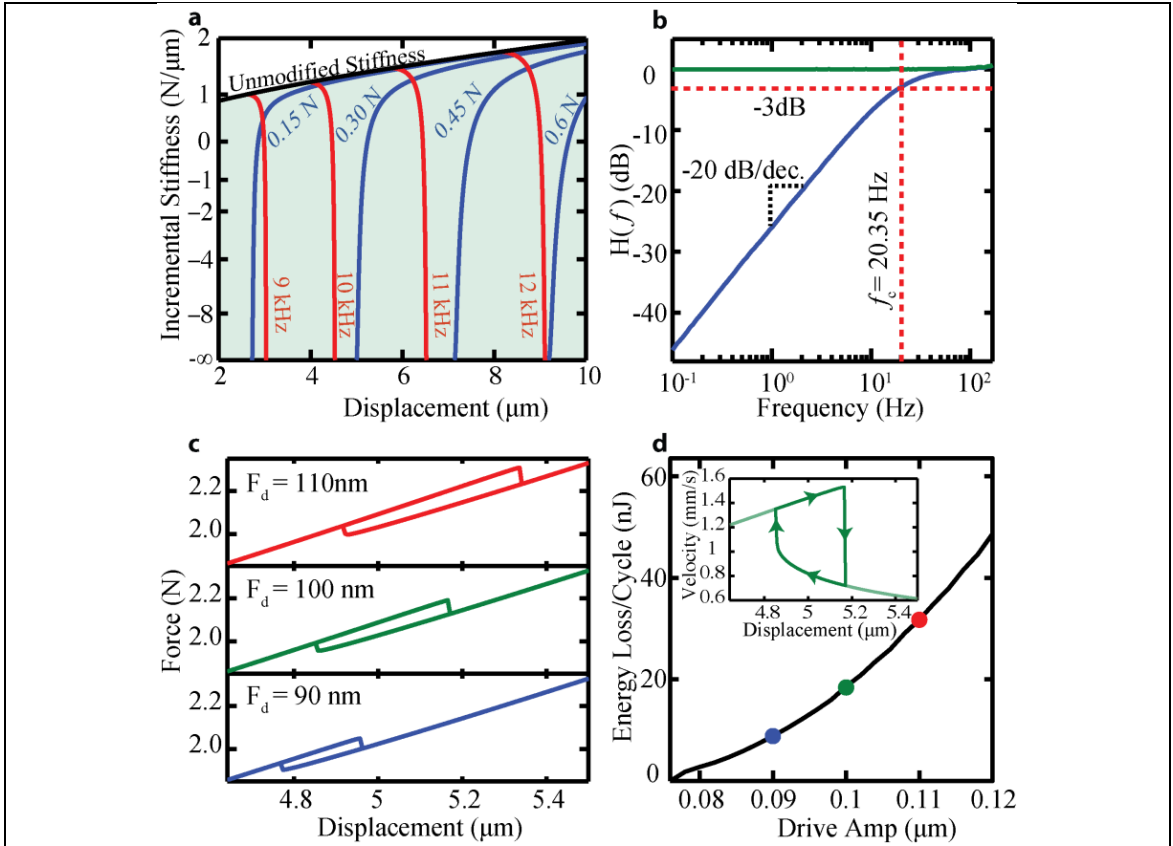

**Figure 4 | Theoretical Investigation. a.** Map relating the excitation parameters with the modified incremental stiffness and displacement point. **b.** Zero frequency band gap obtained by choosing excitation parameters corresponding to zero stiffness for the lattice. The blue and green line show the force transmitted with the defect drive on and off, respectively. The dotted red line shows the band edge frequency, $f_c$. When the drive is on, there is a band gap and when the drive is off the lattice acts as a linear spring. **c.** Force-displacement relationships of the system when it is driven above the bifurcation amplitude. The presence of a tunable hysteresis allows the system to be used as a tunable damper. **d.** As the drive amplitude increases past the bifurcation, the hysteretic losses per cycle increase. The highlighted points correspond to the results in c. The inset in **d** shows the hysteresis in the defect vibration amplitude as the lattice is

> compressed cyclically.

A remarkable feature of the tuning mechanism presented is that it can achieve zero incremental stiffness. In this region the material will support a load, but it will not transmit any vibration to it, which is of great practical relevance [24]. In the zero stiffness region the lattice will have a zero frequency band gap. We simulate this band-gap using our numerical model. In the simulation, we apply a very small amplitude periodic deformation in one end of the chain, and monitor the transmitted force at the other end (Fig 4b). We can see that the band gap exists only at low frequencies, and that that high frequency deformations can propagate without attenuation. We quantify the width of the band gap by fitting the transmission to a first order high pass filter, $H(f) = (f/f_c)/\sqrt{1+(f/f_c)^2}$. This results in a cutoff frequency, $f_c = 20.35\ Hz$. The upper end of the band-gap is a consequence of the fact that the tuned force versus displacement relationship corresponds to the defect mode oscillating in steady-state. When we change the deformation of the lattice, the steady-state oscillation of the defect is perturbed. The system cannot recover the steady state motion instantaneously. The time it takes for the defect mode to relax back to its steady state limits the upper frequency of the band gap. The speed of the system can be analyzed by using a linear perturbation method (Floquet analysis, see Supplementary Information).

At the point where the stiffness reaches minus infinity, the dynamics undergoes a bifurcation. At this bifurcation point the system goes from having a single solution to having multiple stable solutions[22]. This leads to a hysteretic force-displacement response, with the system following different paths when contracting or expanding (Fig. 4c). The area of the hysteresis loop corresponds to the loss of energy incurred as the lattice is driven around a compression cycle. This energy is dissipated by the damping acting on the beads and the excitation force at the defect. Since changing the drive amplitude can control the area enclosed in the hysteresis loop, the system can also act as a tunable damper (Fig. 4d).

This work allows tuning bulk material properties using the excitation of localized defect modes. We anticipate that these results will extend to a variety of materials containing defects. Several questions remain to be answered. Driving multiple defects at different frequencies presents an opportunity to design materials with an arbitrary stress-strain response. Our analytical model shows that by using different nonlinear couplings it is possible to extend the dynamic range of this technique and achieve positive infinite stiffness (see Supplementary Information). In two- and three-dimensional materials this mechanism could lead to controllable anisotropy.

Methods

Numerical Methods. The numerical results are obtained by integrating the Newton's second law of motion for each of the nine particles. The interaction between particles is modelled by the Hertzian contact law[25, 26],

$$F_i(\Delta) = C_i \Delta^{3/2} \; if \; \Delta > 0, \; F_i(\Delta) = 0 \; if \; \Delta \leq 0,$$

in which $\Delta$ the overlap of the Hertzian potential, $C_i$ is a contact factor between particles depending on geometry and material. Using this notation the equations of motion are,

$$m_i \ddot{x}_i = F_i(\delta_i + x_{i-1} - x_i) - F_{i+1}(\delta_i + x_{i-1} - x_i) - m_i \dot{x}_i/\tau,$$

where the force acting on each particle is given by the interaction between neighbors and a linear damping term. The linear damping, $\tau$, as well as the Hertzian contact factors between the lattice and the setup, $C_0$ and $C_{10}$, are fitted experimentally. The equations also contain the masses of each particle, $m_i$, and the equilibrium static overlap between particles $\delta_i$. The sum of the static overlaps is the total displacement (*x*) applied to the lattice. We implement the prescribed displacement boundary condition by adding two additional variables $x_0$ and $x_{10}$ and holding them equal to zero. For experimental fitting and the exact numeric values used in simulations see the Supplementary Information.

**Experimental Methods.** The experiments are carried out on a 1-D lattice of 9 stainless steel 316 beads (McMaster-Carr), with a radius of 9.525 mm. We replace the central bead by a defect bead with a radius of 4.763 mm, and we replace the bead next to the defect by a piezoelectric actuator (Physik Instrumente PD050.31) held between two stainless steel cylinders with a radius of 10 mm and a length of 4 mm. The chain is kept in place using two polycarbonate bars of 6.4 mm radius.

The amplitude of vibration is measured using a laser Doppler vibrometer (Polytec CLV-2534) pointing at the particle next to the defect, and the compression force is measured using a static load cell (Omega LCMFD-50N) and amplified with a gain of 100. The electrical outputs from the sensors are measured with a lock-in amplifier (Zurich Instruments ZI-HF2LI). The strain on the lattice is prescribed using a high-stroke piezoelectric actuator (Physik Instrumente P-841.60). The force-displacement curves are obtained at a steady state by waiting 1.5 seconds to ensure quasistatic behavior.

1. Zhang S, Yin L, Fang N. Focusing Ultrasound with an Acoustic Metamaterial Network. *Physical Review Letters* 2009, **102**(19)**:** 194301.

2. Liu Z, Zhang X, Mao Y, Zhu YY, Yang Z, Chan CT*, et al.* Locally Resonant Sonic Materials. *Science* 2000, **289**(5485)**:** 1734-1736.

3. Pendry JB, Schurig D, Smith DR. Controlling Electromagnetic Fields. *Science* 2006, **312**(5781)**:** 1780-1782.

4. Zhang S, Xia C, Fang N. Broadband Acoustic Cloak for Ultrasound Waves. *Physical Review Letters* 2011, **106**(2)**:** 024301.

5. Jaglinski T, Kochmann D, Stone D, Lakes RS. Composite Materials with Viscoelastic Stiffness Greater Than Diamond. *Science* 2007, **315**(5812)**:** 620-622.

6. Majidi C, Wood RJ. Tunable elastic stiffness with microconfined magnetorheological domains at low magnetic field. *Applied Physics Letters* 2010, **97**(16)**:** -.

7. Florijn B, Coulais C, van Hecke M. Programmable Mechanical Metamaterials. *Phys Rev Lett* 2014, **113**(17)**:** 175503.

8. Nicolaou ZG, Motter AE. Mechanical metamaterials with negative compressibility transitions. *Nat Mater* 2012, **11**(7)**:** 608-613.

9. Fang N, Xi D, Xu J, Ambati M, Srituravanich W, Sun C*, et al.* Ultrasonic metamaterials with negative modulus. *Nat Mater* 2006, **5**(6)**:** 452-456.

10. Dong L, Stone DS, Lakes RS. Broadband viscoelastic spectroscopy measurement of mechanical loss and modulus of polycrystalline BaTiO3 vs. temperature and frequency. *physica status solidi (b)* 2008, **245**(11)**:** 2422-2432.

11. Lakes RS, Lee T, Bersie A, Wang YC. Extreme damping in composite materials with negative-stiffness inclusions. *Nature* 2001, **410**(6828)**:** 565-567.

12. Lapine M, Shadrivov IV, Powell DA, Kivshar YS. Magnetoelastic metamaterials. *Nat Mater* 2012, **11**(1)**:** 30-33.

13. Graeme WM. New metamaterials with macroscopic behavior outside that of continuum elastodynamics. *New Journal of Physics* 2007, **9**(10)**:** 359.

**Acknowledgment** We acknowledge support from the grants US-AFOSR (FA9550-12-1-0332) and Army Research Office MURI grant US ARO (W911NF-09-1-0436).

# Supplementary Information: Extraordinary stiffness tunablility through thermal expansion of nonlinear defect modes

M. Serra Garcia , J. Lydon , and C. Daraio

## Analytic Modeling

The system considered in this paper consists of a chain of particles coupled through an anharmonic interaction potential (Supplementary Fig 1a). In order to get exact results, the motion of all particles needs to be accounted for. However, studying the dynamics of a large number of particles analytically is a difficult problem. In our system we can avoid this complexity by realizing that most of the motion is concentrated around the defect. This is a consequence of the defect mode being highly localized. This localization allows us to capture all of the essential dynamics of the system by considering a single oscillating particle and assuming that other particles in the lattice displace only quasi-statically (Supplementary Fig 1c, d). By using this simplification, we can qualitatively reproduce all of the effects that we have observed experimentally, such as the tuned force-displacement relationship of the lattice. In order to accomplish this, we consider the system at a prescribed total displacement, and then proceed to calculate the amplitude of vibration of the defect, as well as the static force at the boundary.

At each fixed compression value, we model the defect as a point mass $M$, with a dynamic displacement from equilibrium, $u_d$. The defect is subject to a linear damping $F_d = -b\dot{u}_d$ and a periodic excitation force $F(t) = F_e \cos(\omega t)$. As per our model approximation, we consider the neighboring particles to have a constant displacement from equilibrium denoted by $\Delta$. We also assume that the defect motion happens only at the excitation frequency, and is given by $u_d = A\cos(\omega t + \varphi)$. We replace the particles between the defect neighbors and the walls by a linear spring with a force relation $F(\Delta) = F_0 + K_C \Delta$, where $F_0$ is the static force at equilibrium and $K_C$ is calculated by linearizing the interaction force of all the particles after the defect's neighbors and lumping them into a single spring.

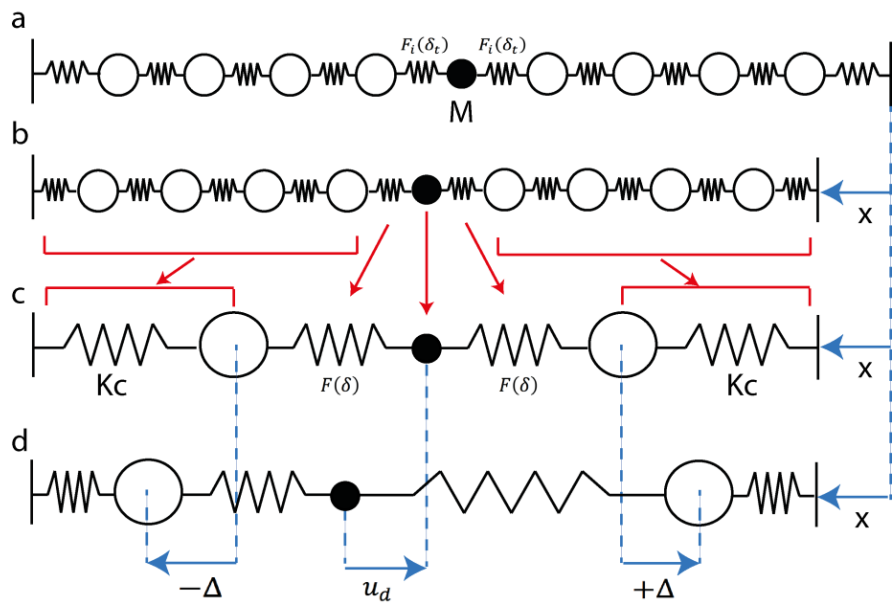

**Supplementary Figure 1 | Analytical modeling of the system**. **a** Initial lattice with no deformation. The lattice consists of a chain of particles, where the central particle is a defect having a mass $M$ smaller than the rest of the particles. The defect interacts with the neighbors through a nonlinear force $F_i(\delta_t)$, where $\delta_t$ is the total distance separating the defect and the neighbors. **b** Deformed lattice. The lattice boundary has been displaced by an amount x. **c** Simplified system used in the analytical approximation. For each fixed displacement value x, the interaction potential between the defect and the neighbors is approximated by a third order polynomial $F(\delta)$, where $\delta = \delta_t - \delta_0$, $\delta_0$

being the equilibrium distance between the defect and the neighbors in the deformed lattice with no defect drive. All the other beads in the chain are approximated by two linear springs $K_C$, with $K_C$ calculated independently for each deformation value x. **d** Simplified system with the defect in motion. The defect is displaced from equilibrium by an amount $u_d$. The two neighboring beads are statically pushed away from it by an amount $\Delta$ due to thermal expansion.

We further simplify the system by performing a Taylor expansion of the nonlinear spring connecting the defect mode with the two half-lattices on each side. We take the Taylor expansion up to third degree, $F(\delta) = F_0 + k\delta + k'\delta^2 + k''\delta^3$. Here, $F_i(\delta_t)$ is approximated by the $F(\delta)$. A force including terms up to third degree is able to capture static equilibrium, linear oscillation, thermal expansion and resonance bending effects. The expansion is calculated around the inter-particle distance at rest, denoted by $\delta_0$. At each deformation value, we calculate the coefficients in the Taylor expansion for the defect-neighbor interaction and the linearized spring constant for the half-lattices. This model results in an equation of motion for the single defect particle and an equation for the static equilibrium of the defect's neighbors. Note that, due to the symmetry of the system, we only need a single equilibrium equation for the two neighbors.

$$M\ddot{u}_d + b\dot{u}_d - F(-u_d - \Delta) + F(u_d - \Delta) = F_e \cos \omega t \quad \text{Eq. S1a}$$

$$K_C \Delta + F_0 = F(u_d - \Delta) \quad \text{Eq. S1b}$$

To solve for the amplitude and static force, we perform a harmonic balance[S1] including only components at the excitation frequency, and discarding terms containing powers of $A^2$ above 3. We neglect higher frequency components because they are significantly lower in the frequency spectrum of the defect's vibration. For the neighbor's equation, we neglect all the harmonic terms and take only the zero-frequency component force. This procedure results in the following condition for the amplitude of the defect:

$$A^2\left[\left(2k+\frac{3}{4}\left[2k''-\frac{8}{3}\left(\frac{k'^2}{K_c+k}\right)\right]A^2-M\omega^2\right)^2+(b\omega)^2\right]-F_e^2=0 \quad \text{Eq. S2}$$

The harmonic balance condition allows us to determine the vibration amplitude of the defect, since all other variables are known: The parameters $k$, $k'$, $k''$ and $K_c$ depend on the total deformation of the lattice, which is prescribed. $F_e$ and $\omega$ describe the defect excitation and are also prescribed. The defect's mass $M$ and damping $b$ are fixed parameters of the system.

We can get further insight on the properties of this system by realizing that the amplitude condition (Eq. S2) is identical to the one that is obtained by performing the same harmonic balance procedure on a Duffing oscillator. A Duffing oscillator is a single degree of freedom dynamical system described by the equation $\ddot{x}+\frac{1}{\tau}\dot{x}+\omega_o^2 x+\alpha x^3 = F_d/M$, and is an extremely well studied system. In order to transform our system into a Duffing oscillator, we use the equations:

$$\omega_0^2=\frac{2k}{M} \quad \text{Eq. S3a}$$

$$\alpha=\frac{1}{M}\left(2k''-\frac{8}{3}\frac{k'^2}{K_c}\right) \quad \text{Eq. S3b}$$

Knowing the vibration amplitude of the defect, it is possible to determine the thermal expansion, and therefore the force at the boundary. To do so we use the defect neighbor's equilibrium equation, and the fact that the force on the linearized spring $K_c$, that connects the defect's neighbors to the boundary, is the same on both ends of the spring.

$$F_b = F_0 + \frac{1}{2}\left(\frac{k'}{1+\frac{k}{k_c}}\right)A^2 \qquad \text{Eq. S4}$$

As expected, the force at the boundary is the sum of the force without any defect drive, and a thermal expansion term that increases with increasing defect motion. The thermal expansion is a consequence of the asymmetric terms in the interaction potential. During a period of the defect oscillation around an equilibrium point, symmetric terms result in an equal amount of attractive and repulsive force. In contrast, asymmetric terms introduce different amounts of attractive and repulsive force, and therefore produce a net effect in the force at the boundary.

Since the analytical model allows us to predict the static force at each displacement value, we can differentiate this prediction with respect to the displacement in order to obtain the stiffness (Eq. S5). This equation contains the original stiffness of the lattice, a term relating changes in force at the boundary to changes in the vibration amplitude of the defect, and a term due to the change in the thermal expansion coefficient as the lattice is compressed.

$$k = \frac{dF_b}{dx} = \frac{dF_0}{dx} + \left(\frac{k'}{1+\frac{k}{K_c}}\right)A\frac{dA}{dx} + \frac{1}{2}A^2\frac{d}{dx}\left(\frac{k'}{1+\frac{k}{K_c}}\right) \qquad \text{Eq. S5}$$

The term $dA/dx$ can be found implicitly from the harmonic balance. This is done by thinking of the balance condition (Eq. S2) as a function of the amplitude and displacement, and noting that the amplitude itself depends on the displacement:

$$\psi(A(x),x) = A^2\left[\left(2k(x) + \frac{3}{4}\left[2k''(d) - \frac{8}{3}\left(\frac{k'(x)^2}{K_c(x)+k(x)}\right)\right]A^2 - M\omega^2\right)^2 + (b\omega)^2\right] - F_e^2 \quad \text{Eq. S6}$$

Since this function must stay constant at zero for all displacements, its total derivative with respect to the displacement must also be zero:

$$\frac{d\psi}{dx} = \frac{\partial \psi}{\partial A}\frac{dA}{dx} + \frac{\partial \psi}{\partial x} = 0 \quad \text{Eq. S7}$$

From the previous equation, it is possible to obtain a closed expression for $dA/dx$, provided that the amplitude of oscillation is known:

$$\frac{dA}{dx} = -\frac{\left(\frac{\partial \psi}{\partial x}\right)_A}{\left(\frac{\partial \psi}{\partial A}\right)_x} \quad \text{Eq. S8}$$

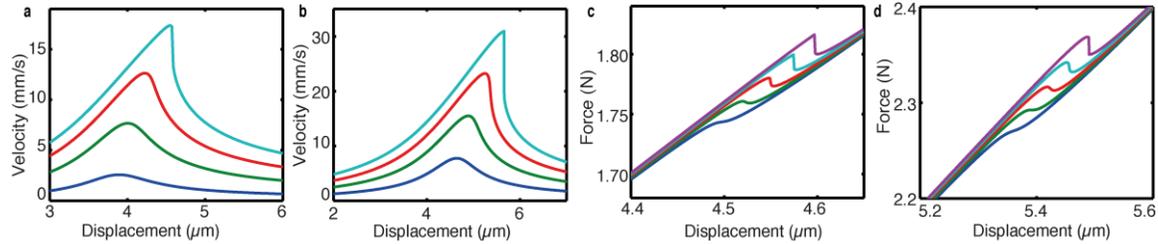

**Supplementary Figure 2 | Analytical predictions and comparison to numerical results**. **a** Defect response obtained by numerically integrating the equations of motion for the full system. **b** Defect response predicted by the analytical model **c** Force- displacement relation of the material obtained through numerical integration **d** Force-displacement relation obtained analytically. All panels are calculated for an excitation frequency of 10.5 KHz at increasing excitation amplitudes.

We have created a simplified model that captures the tuning of the incremental stiffness through the excitation of local defect modes. We use the model to engineer the nonlinear interaction potential so it allows us to tune the stiffness to arbitrarily positive values. This is accomplished by looking at the stiffness equation (Eq. S5). When the changes in the stiffness are very large, the term $k'A\, \partial A/\partial x$ is always dominant. This is because $\partial A/\partial x$ can grow arbitrarily large, while the other terms in the equation are bounded. The term's contribution to the stiffness of the chain, $K$, is given by:

$$\Delta K = -\frac{A}{1+\frac{k}{K_c}\left(\frac{\partial \psi}{\partial A}\right)_x}\frac{k'\left(\frac{\partial \psi}{\partial x}\right)_A}{}$$

This contribution is large when the system approaches a bifurcation. When that happens, $\partial \psi/\partial x$ tends to zero. Depending on the sign of the numerator $-k'(\partial \psi/\partial x)$, the stiffness will grow arbitrarily positive or arbitrarily negative. We study this value for a power law potential, $F = Ax^p$ (Supplementary Fig. 3a). When the exponent p is between 0 and 1, the numerator is positive. In lattices with this kind of interaction force law, the stiffness can be tuned to arbitrarily positive values (Supplementary Figs. 3b and 3c). Recently proposed theoretical work[S2], combined with novel microfabrication techniques[S3] should enable the design of mechanical lattices with tailored interaction potentials. Therefore, it should be possible to create materials with stiffness that can be tuned over a broad range to positive or negative values.
21

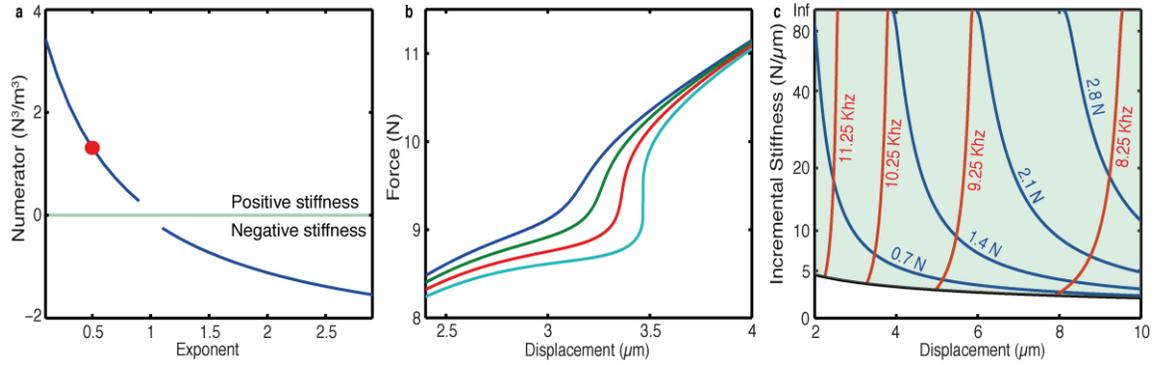

**Supplementary Figure 3 | Stiffness tuning to positive infinity. a** Stiffness numerator corresponding to a power law potential $F = \frac{1}{p}\delta_t^p$ as a function of the exponent. Calculated for a chain of 9 particles with $\delta_T = 1$. Parameters are $M = 1$ and $b = 0.025$. **b** Force-displacement curves for a 9-particle lattice with a power law interaction force exponent $F = Ap^{0.5}$, $A = 5600$. The exponent 0.5 is indicated as a red dot in **a**. The curves correspond to an excitation frequency of 10.5 KHz and increasing excitation forces. **c** Map relating the applied excitation frequency and amplitude to the stiffness, for the same system as **b**. In all panels the interaction force law is assumed to be equal between all neighboring particles. For **b** and **c**, the defect's mass and damping are the same as in Supplementary Fig 2.

## Transient Analysis in Tunable Stiffness

In this system we are driving a defect mode and utilizing the changes in the resonance that occurs as we approach the bifurcation. Essential to this phenomenon of tunable stiffness, is the assumption that system remains at the steady state response. Each time there is an incremental change in the overall displacement of the lattice, there is also a perturbation to the steady state. This means that changes to the incremental stiffness are limited to lower frequencies. So a natural question that arises is, how slow is slow enough?

In the linear model the time that it take for a system to relax back to the steady state solution is dictated by the dissipation, i.e., the quality factor of the system. The perturbation that results from compressing the system decays exponentially. The life of the transient is determined by the system dissipation. However, as the nonlinear system approaches the bifurcation, the slope on one side of the resonance becomes steeper. At the bifurcation point the slope is infinite, and there are two solutions. From a stability point of view, at exactly this point the system is on top of a saddle and does not prefer one solution over the other. Therefore a perturbation takes an infinite amount of time to proceed to the next solution. This means that the time it takes perturbations of the system to relax to the steady state is between the linear dissipation time constant and infinity (which occurs only at the bifurcation point).

The incremental stiffness is limited up to a certain cutoff frequency. From the above qualitative argument when the stiffness is only slightly modified (i.e., in a small amplitude case where the linear assumption holds) the system should react at the speed of the dissipation (quickly). However when the incremental stiffness is being modified more significantly the system is closer to the bifurcation and the perturbations take longer to settle.

Floquet analysis in driven-damped systems is an ideal tool to study the reaction speed of the system. In the context of nonlinear ODE's, Floquet analysis describes a systems relaxation back to a periodic, limit cycle solution. The magnitudes of the Floquet multipliers, $\lambda_i$, of a system describe how linear perturbations to a periodic orbit either grow or decay after a single period, $T$.

$$\boldsymbol{v}(t+T) = \lambda_i e^{i\phi_i} \boldsymbol{v}(t).$$

This equation relates the solution, **v**, of the ODE from one period to the next. The phase, $\phi$, indicates a frequency content of the associated multiplier. When there is dissipation the magnitudes of the multipliers are less than one, $\lambda_i < 1$, indicating that the transients decay. This means that a perturbation decays to the time periodic solution with decay factor, $\lambda$, each period. We can therefore evaluate an effective time constant that depends on how close the system is to the bifurcation,

$$\tau = \frac{T}{ln(\lambda_{max})},$$

where $T$ is the period of the driving frequency and $\lambda$ is the magnitude of the largest Floquet multiplier. It is clear here that as the magnitude of the multiplier approaches one, the denominator goes to zero and the time constant approaches infinity.

As the system approaches the bifurcation (i.e., as the drive amplitude is increased) Floquet multipliers collide along the positive real axis, and one begins to increase in magnitude. The magnitude of the multiplier, or equivalently the time constant, limits the speed at which tuned incremental stiffness can be achieved. As the system reaches the bifurcation, where infinite stiffness occurs, the Floquet multiplier has a magnitude of one and the time constant is undefined.

Supplementary Figure 4 shows the time constants (a), amplitude response (b) and stiffness (c), as the lattice is compressed. The time constant reaches a maximum as the system approaches a bifurcation. This can be seen in the steepening of the amplitude response. This also corresponds to regions of highly modified incremental stiffness. It is important to emphasize that although we are discussing dynamics from the point of view of being slow enough, the system actually reacts quite fast for many practical applications.

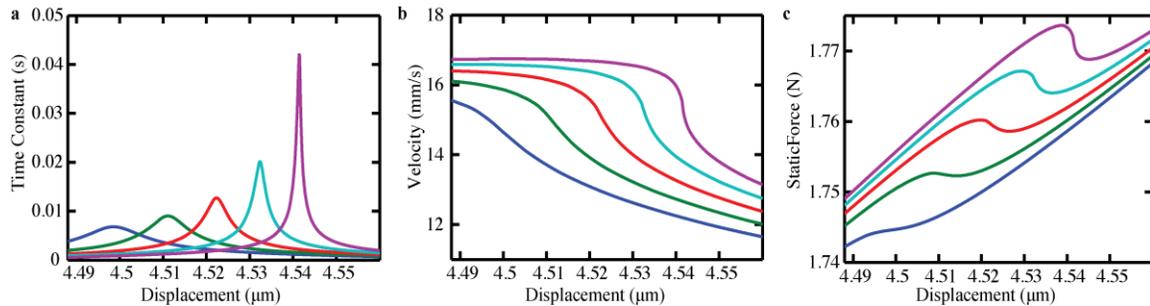

**Supplementary Figure 4 | Transient Analysis**. **a** Time constants that dictate the speed of the relaxation back to steady state. **b** As the system approaches a bifurcation, the amplitude response becomes steeper, and **c** the stiffness is modified more significantly. This is accompanied by longer relaxation time constants, which limits the speed of the system.

In Figure 4b of the paper we show a band gap at zero frequency. The band gap is not due to a local potential, but instead by tuning the stiffness to zero. This discussion indicates that above a certain frequency the system should not be able to react quickly enough to the perturbations applied by the excitation. We calculate the time constant at the excitation required for the zero frequency band gap,

$$\tau_0 = 7.807 \ ms.$$

The time constant that we calculate using Floquet analysis predicts the frequency cutoff remarkably well, $f_\tau = 1/(2\pi\tau) = 20.39$, compared to the frequency cutoff from Figure 4b in the paper which is calculated from fitting to the first order filter, $f_c = 20.35$. The frequency cutoff here is due to a completely different effect than the linear band gap in a periodic crystal. In a traditional band gap, the wave vector becomes imaginary and is reflected. Here, the stiffness of the system is simply zero.

**Model Parameters**

In the numerics, we simplify the complex dynamics of the granular chain to a discrete particle model following[S4]. The table below lists the values used in the simulations, which are either measured experimentally or fit for. We fit for two parameters in our model.

The first is the lattices' Hertzian contact stiffness at the ends. We found that the experimental support expands slightly as a result of its finite stiffness. We include this by modifying the contact stiffness at the edges of the chain. This does not affect the dynamics at the defect site. Supplementary Figure 5 shows the Hertzian fit used to find the total stiffness of the chain and supporting structure, $C_{fit}^{-2/3} = \sum_i C_i^{-2/3}$.

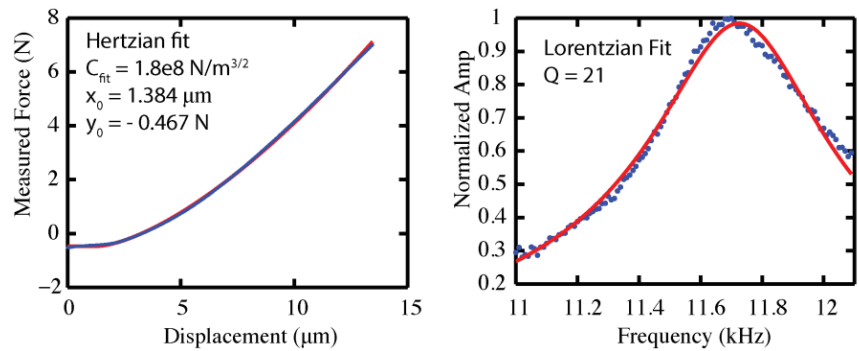

**Supplementary Figure 5 | Experimental fits to determine numerical parameters**. **a** Fit of the static response of the chain to Hertzian Force Law. **b** Fit of the linear amplitude response of the defect to a Lorentzian to determine the linear dissipation of the chain.

The second parameter we fit for is the linear dissipation of the particles. To find this, we perform a frequency sweep at low amplitude drive excitations in the experiment and fit the measured amplitude response to a Lorentzian. We measure the quality factor of the mode and then choose a dissipation time constant for the numerical model that results in the same quality factor for the defect mode.

The contacts at the excitation particle are modified to include the sinusoidal expansion of the chain. Because the stiffness of the piezoelectric disk is much larger than the Hertzian contacts, the piezo expands proportional to the voltage applied, and the assembled structure can be assumed to move as a single expanding bead.

Table 1 – Model Parameters ($C_1$, $C_{10}$, masses, $\tau$ are experimentally measured)

| | |
|---|---|
| $C_i = 9.7576 \ N/\mu m^{3/2}, \quad i \neq 1,5,7,10$ | Sphere – sphere contact stiffness |
| $C_1 = C_{10} = 1.7106 \ N/\mu m^{3/2}, \quad i = 1,10$ | Boundaries contacts stiffness |
| $C_5 = 7.9670 \ N/\mu m^{3/2}$ | Defect sphere contact stiffness |
| $C_7 = 13.799 \ N/\mu m^{3/2}$ | Excitation particle – sphere contact stiffness |
| $m_i = 28.4 \ g, \quad i \neq 6,7$ | Sphere mass |
| $m_6 = 3.6 g, \quad m_7 = 20.2 g$ | Defect mass, and excitation particle mass |
| $\delta_i = \left(\frac{F_0}{C_i}\right)^{2/3}, \quad i \neq 6,7$ | Equilibrium spatial overlap |
| $\delta_i = \left(\frac{F_0}{C_i}\right)^{2/3} + \frac{B}{2}\cos(2\pi f_d t), \quad i = 6,7$ | Spatial overlap including the harmonic signal applied to the excitation particle |
| $\tau = 0.275 ms$ | Dissipation time constant |